\documentclass[pre,twocolumn,amsmath,amssymb,letterpaper,showpacs]{revtex4}

\usepackage{bm}
\usepackage{mathptmx} 

\usepackage{graphicx}

\begin{document}

\title{Calculating free energy profiles in biomolecular systems from fast
  non-equilibrium processes}

\author{Michael Forney}
\affiliation{Department of Physics \& Astronomy, University of Missouri,
  Columbia, MO 65211}

\author{Lorant Janosi}
\affiliation{Department of Physics \& Astronomy, University of Missouri,
  Columbia, MO 65211}

\author{Ioan Kosztin}
\email[Electronic mail: ]{KosztinI@missouri.edu} 
\affiliation{Department of Physics \& Astronomy, University of Missouri,
  Columbia, MO 65211}

\date{\today}

\begin{abstract}
  Often gaining insight into the functioning of biomolecular systems requires to
  follow their dynamics along a microscopic reaction coordinate (RC) on a
  macroscopic time scale, which is beyond the reach of current all atom molecular
  dynamics (MD) simulations. A practical approach to this inherently multiscale
  problem is to model the system as a fictitious overdamped Brownian particle that
  diffuses along the RC in the presence of an effective potential of mean force
  (PMF) due to the rest of the system.
  By employing the recently proposed \textit{FR method} [I. Kosztin et al., J. of
  Chem. Phys. 124, 064106 (2006)], which requires only a small number of fast
  nonequilibrium MD simulations of the system in both forward and time reversed
  directions along the RC, we reconstruct the PMF: (1) of deca-alanine as a
  function of its end-to-end distance, and (2) that guides the motion of potassium
  ions through the gramicidin A channel. In both cases the computed PMFs are found
  to be in good agreement with previous results obtained by different methods.
  Our approach appears to be about one order of magnitude faster than the other
  PMF calculation methods and, in addition, it also provides the position
  dependent diffusion coefficient along the RC. Thus, the obtained PMF and
  diffusion coefficient can be used in a suitable stochastic model to estimate
  important characteristics of the studied systems, e.g., the mean folding time of
  the stretched deca-alanine and the mean diffusion time of the potassium ion
  through gramicidin A.
\end{abstract}

\maketitle

\section{Introduction}
\label{sec:intro}

The study of the structure-function relationship of biomolecular systems often
requires to follow their dynamics with almost atomic spatial resolution on a
macroscopic time scale, which is beyond the reach of current all atom molecular
dynamics (MD) simulations.
A typical example is molecular and ion transport through channel proteins
\cite{roux02-182}. Indeed, in order to determine the forces that guide the
diffusion of molecules across the channel one needs to know with atomic precision
the structure of the channel protein-lipid-solvent environment. However, the
duration of the permeation process across the channel occurs on a time scale
(e.g., $\mu$s to ms) that may exceed by several orders of magnitude the time scale
of several tens of nanoseconds currently attainable by all atom molecular dynamics
(MD) simulations\cite{CBB01}.
Whenever the dynamic properties of interest of such a system can be described in
terms of a small number of reaction coordinates (RCs) then a practical approach to
this inherently multiscale problem is to model the system as fictitious overdamped
Brownian particles that diffuse along the RCs in the presence of an effective
potential of mean force (PMF) that describes their interaction with the rest of
the system. 

Recently we have proposed an efficient method for calculating simultaneously both
the PMF, $U(R)$, and the corresponding diffusion coefficient, $D$, along a RC,
$R$, by employing a small number of fast nonequilibrium MD simulations in both
forward (F) and time reversed (R) directions \cite{kosztin06-064106}.
The efficiency of this method, referred to as the \textit{FR method}, was
demonstrated by calculating the PMF and the diffusion coefficient of single-file
water molecules in single-walled carbon nanotubes \cite{kosztin06-064106}. The
obtained results were found to be in very good agreement with the results from
other PMF calculation methods, e.g., umbrella sampling
\cite{frenkel2002,roux95-275,torrie77-187}.

To further test its viability, in this paper we apply the FR method to investigate
the energetics of two well-studied exemplary systems, i.e., (i) the helix-to-coil
transition of deca-alanine in vacuum, and (ii) the transport of $K^{+}$ ions in
the gramicidin A (gA) channel protein, inserted in a fully solvated POPE lipid
bilayer.  In each case we seek to calculate the PMF as a function of a proper RC,
i.e., the end-to-end distance ($R$) of deca-alanine and the position
($z$-coordinate) of the potassium ion along the axis of the gA channel.
The computed PMFs are found to be in good agreement with previous results obtained
by using either the Jarzynski equality \cite{park04-5946,park03-3559} or the
umbrella sampling method \cite{frenkel2002,roux95-275,torrie77-187}. However,
compared to these PMF calculation methods our approach is about one order of
magnitude faster and, in addition, also provides the position dependent diffusion
coefficient along the RC. 
Thus, by employing the computed PMF and diffusion coefficient in a suitable
stochastic model we could estimate important characteristics of the studied
systems, e.g., the mean folding time of the stretched deca-alanine and the mean
first passage time of $K^{+}$ through the gA channel. 

The remaining of the paper is organized as follows. To make the presentation
self-contained, in Sec.~\ref{sec:theory} a brief description of the FR method is
provided, along with the theory used to analyse our results. The study of
deca-alanine is described in Sec.~\ref{sec:10ala}, while that of $K^{+}$ transport
in the gA channel in Sec.~\ref{sec:gA}. Finally, Sec.~\ref{sec:conc} is reserved
for conclusions.

\section{Theory}
\label{sec:theory}

By definition, for a classical mechanical system described by the Hamiltonian
$H_0(\Gamma)$, the PMF (Landau free energy), $U(R)$, along a properly chosen RC
($R$) is determined from the equilibrium distribution function of the system by
integrating out all degrees of freedom except $R$, i.e., \cite{frenkel2002}
\begin{equation}
  \label{eq:pmf1}
  e^{-\beta U(R)}\equiv p_0(R)=\int d\Gamma \frac{e^{-\beta H_0(\Gamma)}}{Z_0}
  \delta[R-\tilde{R}(\Gamma)] \;.
\end{equation}
Here $p_0(R)$ is the equilibrium distribution function of the RC, $Z_0$ is the
partition function, $\beta=1/k_BT$ is the usual thermal factor, and $\delta(R)$ is
the Dirac-delta function whose filtering property guarantees that the integrand in
Eq.\eqref{eq:pmf1} is nonzero only when $\tilde{R}(\Gamma)=R$. In this paper we
use the convention that $R$ [or $R(t)$] is the target value, while
$\tilde{R}\equiv \tilde{R}(\Gamma)$ is the actual value of the RC.
Also, it is convenient to use $k_BT$ as energy unit. Thus, in Eq.~\eqref{eq:pmf1}
one needs to set $\beta=1$.

Unfortunately, by using equilibrium MD simulations the direct application of
Eq.~\eqref{eq:pmf1} is practical only for calculating $U(R)$ about its local
minimum. 
An efficient way to properly sample $R$ is provided by \emph{steered molecular
  dynamics} (SMD) \cite{isralewitz01-13} in which the system is guided, according
to a predefined protocol, along the RC by using, e.g., a \emph{harmonic guiding
  potential}
\begin{equation}
  \label{eq:hgp}
  V_R(\tilde{R}(\Gamma)) = \frac{k}{2}[\tilde{R}(\Gamma)-R]^2 \;,
\end{equation}
where $k$ is the elastic constant of the harmonic guiding potential. With this
extra potential energy, the Hamiltonian of the new biased system becomes $H_R =
H_0 + V_R(\tilde{R})$.  As a result, atom ``$j$'' in the selection that define the
reaction coordinate will experience an additional force
\begin{equation}
  \label{eq:hgp2}
\mathbf{F}_j =
-\frac{\partial{V_R}}{\partial{\mathbf{r}_j}}=-k  [\tilde{R}(\Gamma)-R]
\frac{\partial{\tilde{R}(\Gamma)}}{\partial{\mathbf{r}_j}}\;. 
\end{equation}
By choosing a sufficiently large value for the elastic constant $k$, i.e., the
so-called \emph{stiff-spring} approximation \cite{jensen02-6731,park04-5946}, the
distance between the target and actual value of the RC at a given time can be kept
below a desired value.

In constant velocity SMD simulations \cite{isralewitz01-13}, starting from an
equilibrium state characterized by $R(0)$, the target value of the RC (or control
parameter) $R(t)$ is varied in time according to $R(t)=R(0)+vt$, $0\le t\le\tau$,
where $v$ is the constant pulling speed. For each such forward (F) path there is a
time reversed (R) one in which the system starts from an equilibrium state
corresponding to $R(\tau)$ and reaches $R(0)$ according to the protocol
$R_R(t)=R_F(\tau-t)=R(\tau)-vt$, $0\le t\le\tau$.
The external work done during a SMD simulation is given by
\begin{equation}
  \label{eq:work}
W_F = \int_{R_0}^{R(t)} dR
\left[ \partial{V_R(\tilde{R})}/\partial{R}\right] = k\int_{R_0}^{R(t)}
dR (R-\tilde{R})\;. 
\end{equation}
The F and R work distributions are not independent but related through the
\emph{Crooks Fluctuation Theorem} \cite{crooks00-2361}
\begin{equation}
  \label{eq:tft}
  \frac{P_F(W)}{P_R(-W)}=e^{W_{dF}}\;,
\end{equation}
where the F dissipative work is given by
  \begin{equation}
    \label{eq:wdF}
    W_{d\,F}=W_{F}-\Delta{U}\;,
  \end{equation}
with $\Delta{U}=U[{R(\tau)}]-U[{R(0)}]$.
In principle, the PMF can be determined from the so-called Jarzynski equality (JE)
\cite{hummer01-3658} 
\begin{equation}
  \label{eq:JE1}
  \left\langle \exp(-W_{dF}) \right\rangle  = 1\;,
\end{equation}
that follows directly from Eq.~\eqref{eq:tft} \cite{crooks00-2361}. Within the
stiff-spring approximation the sought PMF is given by the second cumulant
approximation \cite{park04-5946,park03-3559,jensen02-6731}
\begin{eqnarray}
  \label{eq:pmf-smd-2}
  \Delta{U_F(R)}&=&-\log \left\langle \exp[-W_F(R)] \right\rangle \approx \langle{W_F}\rangle
   -\sigma^2_F /2\;,\\
  \sigma^2_F &=&  \langle{W_F^2}\rangle - \langle W_F\rangle^2\;,\nonumber
\end{eqnarray}
where $\sigma^2_F$ is the variance (2nd cumulant) of the F work.
Also, within the stiff-spring approximation the work distribution function
$P_F(W)$ is Gaussian and, therefore, the cumulant approximation
\eqref{eq:pmf-smd-2} is exact \cite{park04-5946}. However, in practice
Eq.~\eqref{eq:pmf-smd-2} is valid only close to equilibrium because SMD pulling
paths can sample only a narrow region about the peak of the Gaussian $P_F(W)$,
while the validity of JE is crucially dependent on very rare trajectories with
negative dissipative work ($W_d < 0$).  Thus, in general, having only a few SMD
trajectories one can determine fairly accurately the mean work $\left\langle W_F
\right\rangle$ but not the variance $\sigma_F^2$, which in most cases is seriously
underestimated.

In the FR method this shortcoming is eliminated by combining both F and R pulling
trajectories and employing Eq.~\eqref{eq:tft}, which is more general than the JE
\eqref{eq:JE1}. Within the stiff-spring approximation, Eq.~\eqref{eq:tft} implies
that the F and R work distribution functions are identical but displaced
Gaussians, and the PMF and the mean dissipative work $W_d\equiv W_{d\,F}=W_{d\,R}$
can be determined from the following simple equations \cite{kosztin06-064106}
\begin{subequations}
\begin{equation}
  \label{eq:fr-2b}
  \Delta{U}=(\langle{W}_F\rangle -\langle{W}_R\rangle)/2\;,
\end{equation}
\begin{equation}
  \label{eq:fr-2c}
  \langle{W}_d\rangle = (\langle{W}_F\rangle +\langle{W}_R\rangle)/2\;,
\end{equation}
and
\label{eq:fr-2}
  \begin{equation}
    \label{eq:fr-2a}
    \sigma^2\equiv\sigma_F^{2}=\sigma_R^2 = 2\langle{W}_d\rangle\;.
  \end{equation}
\end{subequations}
Equations \eqref{eq:fr-2} are the key formulas of our FR method for calculating PMFs
from fast F and R SMD pullings. 
Clearly, the superiority of the FR method, for calculating the PMF (and the mean
dissipative work), compared to the one based on the JE equation is due to the fact
that Eqs.\eqref{eq:fr-2} contain only the mean F and R work (whose values can be
estimated rather accurately even from a few SMD trajectories) and not the
corresponding variance. In fact the latter (see Eq.~\eqref{eq:fr-2a}) is also
determined by the mean F and R work.

Although, strictly speaking, the FR method can only determine the PMF difference
between initially equilibrated states connected by F and R SMD trajectories, in
practice we find that in many cases Eqs~\eqref{eq:fr-2b}-\eqref{eq:fr-2c} give
good results even between the division points $R_i$, $i=1,\ldots,N-1$, of the
interested interval [$R_0=R(0)$,$R_N=R(\tau)$]. 
The reason for this is that for a stiff harmonic guiding potential the equilibrium
distribution of the RC is a narrow Gaussian that can be sampled through very short
MD simulations. Thus, even if the system is far from equilibrium due to fast
pulling by a sufficiently stiff spring, the instantaneous value of the RC will
always be sufficiently close to its equilibrium value. However, even in such cases
the pulling speed should not exceed values that would cause excessive perturbation
to the rest of the degrees of freedom of the system.
Thus, the number of division points, $N$, does not need to be large, implying a
fairly small computational overhead for the equilibration of the system at $R_i$,
$i=1,\ldots,N-1$.

An alternative approach for calculating the PMF difference between two equilibrium
states connected by $n_F$ forward and $n_R$ reverse SMD paths is based on the
\emph{maximum likelihood estimator} (MLE) method applied to Crooks' fluctuation theorem
\eqref{eq:tft} \cite{shirts03-140601}, i.e.,
\begin{eqnarray}
  \label{eq:ar-2}
    &&\sum_{i=1}^{n_F}\frac{1}{1+n_F/n_R \exp(W_{F i}-\Delta{U})}\\
    && -
    \sum_{i=1}^{n_R}\frac{1}{1+n_R/n_F \exp(-W_{R i}-\Delta{U})} = 0 \;.\nonumber
\end{eqnarray}
We use Eq.~\eqref{eq:ar-2} to test the accuracy of the PMF results obtained with
our FR method.

Finally, since it is reasonable to assume that $\overline{W}_d$ is proportional to
the pulling speed $v$, one can readily determine the position dependent friction
coefficient $\gamma(R)$ from the slope of the mean dissipative work $\gamma(R)
=\left( d\langle{W}_d(R)\rangle/dR \right)/v$. Then, the corresponding diffusion
coefficient is given by the Einstein relation (in $k_BT$ energy units)
\cite{kosztin06-064106} 
\begin{equation}
  \label{eq:fr-diff}
  D(R) = \gamma(R)^{-1}=v \left( d\langle{W}_d(R)\rangle/dR \right)^{-1} \;.
\end{equation}
Once both $U(R)$ and $D(R)$ are determined, the dynamics of the reaction
coordinate on a macroscopic time scale can be described by the Langevin equation
corresponding to an overdamped Brownian particle\cite{zwanzig2001}
\begin{subequations}
\label{eq:fr-L-FP}
  \begin{equation}
    \label{eq:fr-L}
    \gamma(R) \dot{R} = -dU(R)/dR + \xi(t)\;,
  \end{equation}
or equivalently, by the corresponding Fokker-Planck equation for the probability
distribution function $p(R,t)$ of the reaction coordinate 
  \begin{equation}
    \label{eq:fr-FP}
    \begin{split}
    \partial_t p(R,t) &= -\partial_R j(R,t)\\ 
    &= \partial_R [D(R) \partial_R p(R,t)]+\partial_R[U'(R) p(R,t)]\;,
    \end{split}
  \end{equation}
\end{subequations}
where $\xi(t)$ is the Langevin force (modeled as a Gaussian white noise) and
$j(R,t)$ is the probability current density.

For example, Eq.~\eqref{eq:fr-FP} can be used to calculate the mean folding time
of deca-alanine (see Sec.~\ref{sec:10ala-results}) from the completely stretched
(coil) conformation $R_c=33~\text{\AA}$ to the folded (helical) conformation
$R_0=14.5~\text{\AA}$ as the corresponding \textit{mean first passage time} (MFPT)
\cite{risken96}, i.e.,
\begin{equation}
  \label{eq:mfpt}
  \tau = \int_{R_c}^{R_0} dR\, e^{U(R)}/D(R) \int_{R_c}^R dR' e^{U(R')}\;.
\end{equation}

\section{Stretching deca-alanine}
\label{sec:10ala}

\begin{figure}
  \centering
  \includegraphics[width=3.4in]{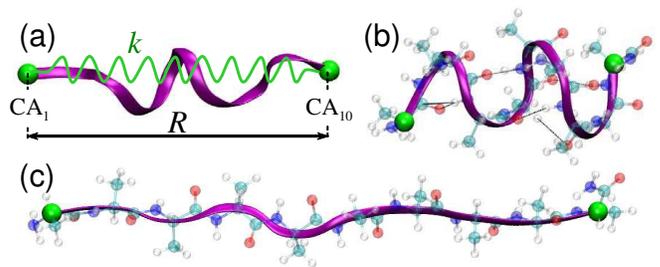}  
  \caption{(color online). (a) Cartoon representation of deca-alanine. The
    reaction coordinate $R$ is defined as the distance between the first (CA$_1$)
    and last (CA$_{10}$) $C_{\alpha}$ atoms, i.e., the end-to-end distance of the
    peptide. The spring, with elastic constant $k$, connecting CA$_1$ and
    CA$_{10}$ corresponds to an elastic guiding potential
    $V(R;t)=(k/2)[R-R_0(t)]^2$ that can be used to cycle deca-alanine between the
    (b) folded and (c) unfolded (completely stretched) conformations. In (b) and
    (c) the backbone (sidechain) atoms are shown in cartoon (CPK)
    representation. In the folded (b) configuration the hydrogen bonds that
    stabilize the $\alpha$-helix are also shown. (Snapshots rendered with the
    program VMD \cite{HUMP96}).}
  \label{fig:1}
\end{figure}

Deca-alanine is a small oligopeptide composed of ten alanine residues
(Fig.~\ref{fig:1}). The equilibrium conformation of deca-alanine, in the absence
of solvent and coupled to an artificial heat bath at room temperature, is an
$\alpha-$helix. The system can be stretched to an extended (coil) conformation by
applying an external force that pulls its ends apart. Once the stretched system is
released it will refold spontaneously into its native $\alpha-$helical
conformation. Thus, this can be regarded as a simple protein unfolding and
refolding problem that can be comfortably studied via SMD simulations due to the
relatively small (104 atoms) system size. It is natural to define the reaction
coordinate as the distance $R$ between the first (CA$_1$) and the last (CA$_{10}$)
$C_{\alpha}$ atoms. To calculate the PMF, $U(R)$, that describes the energetics of
the folding/unfolding process we have use SMD simulations to generate a small
number (in general 10) F and R pulling trajectories and apply the PMF calculation
methods described in Sec.~\ref{sec:theory}, i.e., the FR method
[Eqs.~\eqref{eq:fr-2}], the JE method [Eq.~\eqref{eq:pmf-smd-2}] and the MLE
method [Eq.~\eqref{eq:ar-2}]. The SMD harmonic guiding potential \eqref{eq:hgp}
corresponded to an ideal spring of tunable undeformed length $R(t)$ inserted
between CA$_1$ and CA$_{10}$ (see Fig.~\ref{fig:1}a). Note that this choice of the
guiding potential is more natural than the one customarily used in the literature
in which the atom attached to one of the two ends of the spring is fixed
\cite{park03-3559,henin04-2904,procacci06-164101}.

\subsection{Computer modeling and SMD simulations}
\label{sec:10ala-modeling}

The computer model of deca-alanine was built by employing the molecular modeling
software VMD \cite{HUMP96}.  All simulations were performed with NAMD~2.5
\cite{phillips05-1781} and the CHARMM27 force field for proteins
\cite{MACK92short,MACK98short}.  A cutoff of 12\AA\ (switching function starting
at 10\AA) for van der Waals interactions were used.  An integration time step of
$2$~fs was employed by using the SHAKE constraint on all hydrogen atoms
\cite{miyamoto92-952}.  The temperature was kept constant (at $300$~K) by coupling
the system to a Langevin heat bath.  The system was subjected to several
equilibrium MD and non-equilibrium SMD simulations.  We divided the reaction
coordinate $R\in[13,33]~\text{\AA}$ into ten equidistant intervals (windows)
delimited by the points $R_i=(13+2i)$~\AA, $i=0,\ldots,10$.  Next, a pool of
equilibrium states were generated for each $R_i$ from $4$~ns long equilibrium MD
trajectories. These states were used as starting configurations for the SMD F and
R pulls on each of the ten intervals.  The spring constant in these equilibrium MD
simulations was $k=50$~kcal/mol/\AA$^2$.
The equilibrium length of the folded deca-alanine was determined from two free MD
simulations starting from a compressed ($R=13~\AA$) and the completely stretched
($R=33~\AA$) configurations of deca-alanine. Both simulations led to the same
equilibrium length $R_{eq}=14.5~\text{\AA}$.

In order to calculate $U(R)$ a total of six sets of F and R SMD simulations were
carried out.
In each of the first three sets of SMD runs we used ten
simulation windows, but three different pulling speeds: $v_1=1$~\AA/ps,
$v_2=10^{-1}$~\AA/ps, and $v_0=10^{-4}$~\AA /ps. 
The sets corresponding to $v_{1,2}$ consisted of $10$ F and $10$ R SMD
trajectories. For the quasi-equilibrium pulling speed $v_0$ only one F and R runs
were performed.
In the last three sets of SMD simulations we used a single simulation window,
covering the entire range of the RC, and used the same three pulling speeds as in
the previous SMD runs.
For all six sets of SMD simulations, the stiff-spring constant was
$k=500$~kcal/mol/\AA$^2$.  

To construct the forward and reverse work distribution functions on the segment
$R\in[17,21]~\text{\AA}$, we performed $2000$ F and the same number of R SMD
simulations. In order to generate a sufficient number of starting equilibrium
configurations it was necessary to extend the equilibration runs at both
$R=17$~\AA\ and $R=21$~\AA\ to $5$~ns. In all these simulations we used a pulling
speed of $v=1$~\AA/ps and a spring constant of $k=500$~kcal/mol/\AA$^2$.

Finally, to estimate the mean refolding time of the completely stretched
deca-alanine we performed 100 free MD simulations starting from an equilibrium
configuration corresponding to $R=33$~\AA. As soon as deca-alanine reached its
folded, equilibrium length $R_{eq}=14.5$~\AA\ the simulation was stopped and the
refolding time recorded.

\subsection{Results and Discussion}
\label{sec:10ala-results}

\begin{figure}
  \centering
  \includegraphics[width=3.4in]{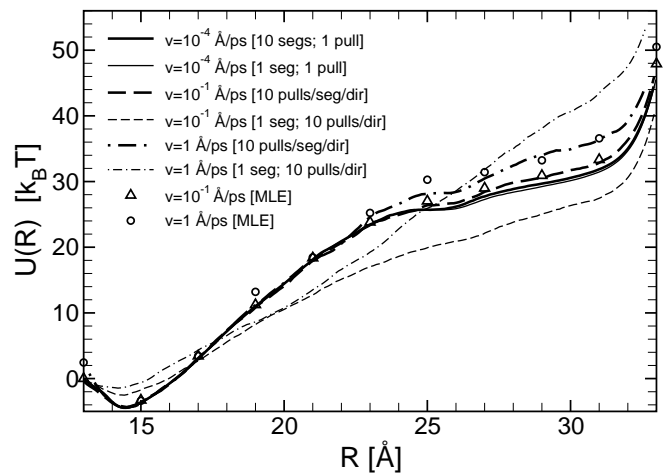}  
  \caption{Potential of mean force (PMF) of deca-alanine as a function of the
    reaction coordinate $R$. The different curves were obtained with the FR
    method by employing different simulation and PMF calculation protocols
    described in the text.}
  \label{fig:2}
\end{figure}

\begin{figure}
  \centering
  \includegraphics[width=3.4in]{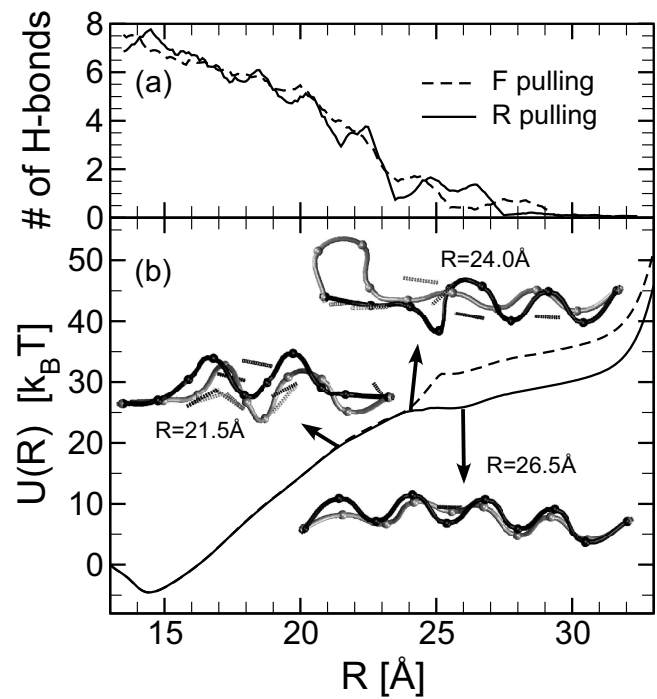}  
  \caption{(a) Variation of the number of hydrogen bonds in deca-alanine during
    the quasi-equilibrium ($v=10^{-4}~\text{\AA/ns}$) F and R pullings. (b) The
    PMF $U(R)$ calculated as the external work done during the quasi-equilibrium F
    (dashed line) and R (solid line) pullings. The discrepancy between the two
    PMFs is most likely due to the difference on how the H-bonds are formed and
    destroyed during the forced folding and unfolding processes, respectively, as
    indicated in the inset snapshots of the peptide. Dark (lite) color corresponds
    to the R (F) process.}
  \label{fig:5}
\end{figure}

The PMFs calculated using the FR method corresponding to the six different pulling
protocols described in Sec.~\ref{sec:10ala-modeling} are shown in
Fig.~\ref{fig:2}. As expected, for the very small pulling speed $v_0$ the system
is in quasi-equilibrium throughout the SMD runs leading to the same (true) PMF
regardless of the number of simulation windows considered. However, while the
dissipative work is negligible for both F and R processes, repetition of these
simulations resulted in different PMFs for $R>24~\text{\AA}$, and it will be
discussed below (see also Fig.~\ref{fig:5}).
Not surprisingly, in case of the very fast pulling speed $v_1$, the PMF for the
single simulation window is rather poor along $R$ except at the end-points of the
window. Indeed, the FR method allows to calculate the PMF difference between two
equilibrium states connected by fast F and R SMD processes that follow the same
protocol. However, it is remarkable that using ten simulation windows, even at
this large pulling speed the resulting PMF is rather close to the real one. For
the still fast pulling speed $v_2$ the situation is similar. While the single
simulation window case lead to a rather poor PMF (though somewhat better than in
the $v_1$ case), the ten simulation windows result is almost indistinguishable
from the true PMF.
For comparison, the PMFs calculated at $R_i$, $i=1,\ldots,10$ using the MLE method
for both $v_1$ and $v_2$ are also shown in Fig.~\ref{fig:2}. 
Based on these results one may conclude that the FR method gives very good PMF
even for fast pulling speeds and using only a few F and R trajectories, provided
that a sufficient number of simulation windows are used.

\begin{figure}
  \centering
  \includegraphics[width=3.4in]{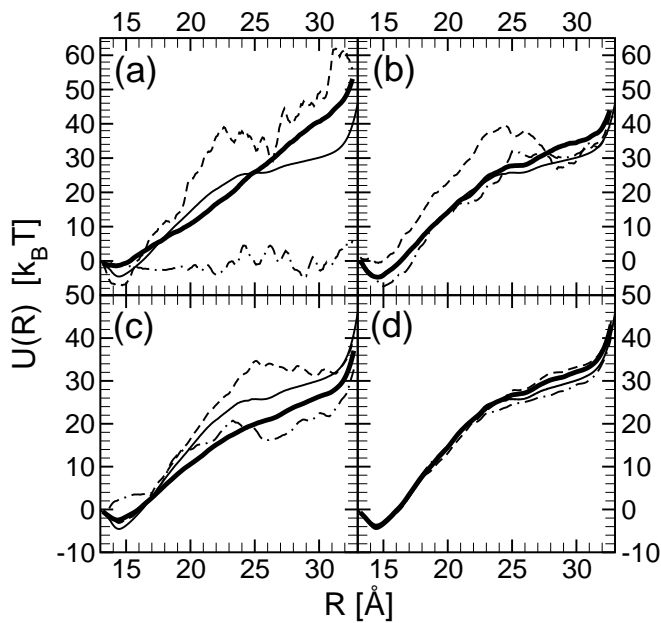}  
  \caption{Comparison between the PMFs $U(R)$ obtained using the FR method
    (thick solid line) and the cumulant approximation of the JE corresponding to
    the ten forward (dashed line) and reverse (dot-dashed line) SMD trajectories,
    respectively. The thin solid line corresponds to the exact PMF. The upper
    (lower) panels correspond to a uniform pulling speed of $1~\text{\AA/ns}$
    ($0.1~\text{\AA/ns}$). The PMFs in the right panels were determined by
    dividing the $20~\text{\AA}$ pulling distance into ten equidistant segments
    (the system being equilibrated in each of the end points of the individual
    segments), while the PMFs in the left panels were determined by considering
    the entire pulling distance as a single segment.}
  \label{fig:3}
\end{figure}

A comparison between $U(R)$ obtained from the FR method and the cumulant
approximation of the JE method (applied separately for the F and for the R SMD
trajectories) are shown in Fig.~\ref{fig:3}. In general, the FR method yields
better PMF in all cases, and especially when one employs (i) one simulation window
(Fig.~\ref{fig:3}a and c), and (ii) a very large pulling speed $v_1$
(Fig.~\ref{fig:3} a and b).
For the ten simulation windows with pulling speed $v_2$ (Fig.~\ref{fig:3} d) the
FR and JE methods are comparable though even in this case the JE F (R) method
systematically over (under) estimates the PMF. Note, however, that an average of
the JE PMFs for the F and R trajectories leads to a result very close to the FR
one.

\begin{figure}
  \centering
  \includegraphics[width=3.4in]{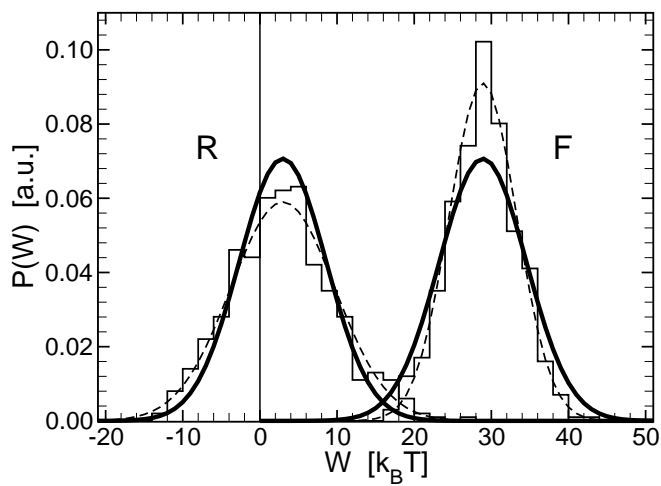}  
  \caption{Histogram of the distribution functions (thin solid lines) of the
    forward ($W_F$) and reverse ($W_R$) works along the segment $R\in
    [17,21]\text{\AA}$. Although the histograms seem to be Gaussian (dashed lines)
    they are not identical as predicted by the FR method (see text for details).}
  \label{fig:4}
\end{figure}

An important prediction of the FR method is that, provided that the stiff-spring
approximation holds, the F and R work distributions are identical Gaussians
centered about the mean F and R work, and therefore shifted by $2\Delta U$. To
test this prediction we have determined the work distribution histogram
corresponding to $2000$ F and a same number of R SMD trajectories corresponding to
the RC segment $R\in[17,21]~\text{\AA}$. The results are shown in
Fig.~\ref{fig:4}. Although the histograms seem to be Gaussian (dashed lines) they
are not identical as predicted by the FR method. In a previous study
\cite{procacci06-164101} the clear deviation from Gaussian of the external work
distribution in case of deca-alanine was pointed out and it was attributed to the
non-Markovian nature of the underlying dynamics of the system. However, in our
case both work distributions look Gaussian and the relatively small but clearly
noticeable difference between them may be due either to the failure of the
stiff-spring approximation or to incomplete sampling. After all, the end-to-end
distance is a poor and insufficient reaction coordinate for describing the folding
and unfolding processes of a polypeptide.

This last point becomes rather clear when the system is subjected to repeated
folding (R) and unfolding (F) processes at the quasi-equilibrium speed $v_0$. At
this speed the system is at almost equilibrium throughout the SMD pulls and one
expect that the PMF is given by the external work, i.e., the dissipated energy
(which is a stochastic quantity) is negligible. While for $R<24~\text{\AA}$ one
gets systematically the same PMF, for $R\in [24,33]~\text{\AA}$ one obtains
different PMFs depending on the direction of pulling, as one can see in
Fig.~\ref{fig:5}b.
A careful inspection of these trajectories reveal that the folding and unfolding
processes occur through different pathways in the above mentioned range of the RC.
Thus, it appears that $R$ is not sufficient to specify the metastable intermediate
states of the system, and a more complete description requires the introduction of
extra order parameters, e.g., the distribution of the hydrogen bonds (H-bonds) in
the peptide. Indeed, the dynamics of the formation and rupture of the H-bonds
during folding and unfolding, respectively, may be rather different. As shown in
the inset snapshots in Fig.~\ref{fig:5}b, the formation of the six H-bonds during
the R process is much more homogeneous than their rupture during the corresponding
F process. This observation is reinforced by the time dependence of the average
number of H-bonds in deca-alanine shown in Fig.~\ref{fig:5}a. Thus there are at
least two distinctive pathways in the helix-to-coil transition of deca-alanine,
both being explored during quasi-static pullings. During fast pulling, however,
one of the pathways is preferred compared to the other.

Finally, as an application of the determined PMF and the diffusion coefficient,
which was found to be approximately constant $D\approx
0.27~\text{\AA}^2/\text{ps}$, we calculated the mean folding time (i.e.,
coil-to-helix transition) by employing Eq.~\eqref{eq:mfpt}. The theoretical result
of $\tau\approx 140$~ps compares rather well with the MFPT of $\approx 100$~ps
obtained from the $100$ free MD refolding simulations described in
Sec.~\ref{sec:10ala-modeling}.

\section{K$^{+}$ transport in gramicidin A channel}
\label{sec:gA}

\begin{figure}[htbp]
   \centering
   \includegraphics[width=3.4in]{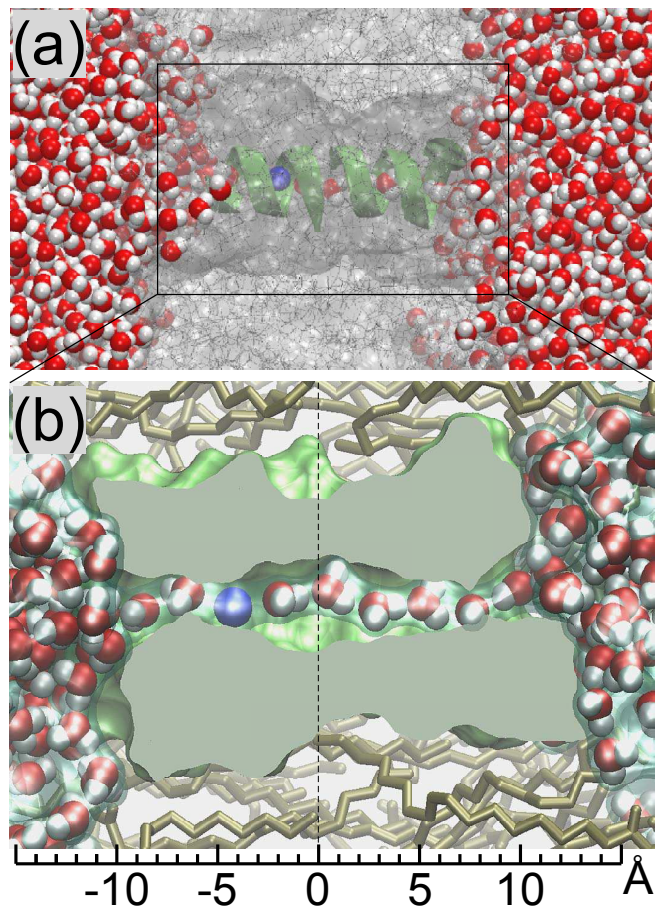} 
   \caption{(color online). (a) Gramicidin-A channel ($\beta$-helix dimer colored
     in green) in POPE lipid bilayer (grey), solvated in water (van der Waals
     representation).  The potassium ions is shown as a blue sphere. (b)
     Cross-section of the gramicidin A channel. The $K^+$ ion (blue) and the water
     molecules move single file inside the pore. (Created with the program VMD
     \cite{HUMP96}).}
   \label{fig:gram-a}
\end{figure}

Gramicidin A (gA) is the smallest known ion channel that selectively conducts
cations across lipid bilayers \cite{wallace86-295}. gA is a dimer of two
barrel-like $\beta$-helices that form a $\sim 26$~\AA\ long and $4-5$~\AA\ wide
cylindrical pore through the lipid membrane (Fig.~\ref{fig:gram-a}). Each helix
consists of 15 alternating Asp and Leu amino acids. Due to its structural
simplicity, gA is an important testing system for ion permeation models, and it
has been extensively studied in the literature both experimentally and through
computer modeling.
NMR studies have shown that each end of the channel has a cation binding site that
is occupied as the ion concentration is increased \cite{tian99-1993} . The
conductance is at maximum when the average ion population in the channel is
one. The backbone carbonyls inside the pore are oriented such that the
electronegative oxygen atoms face inward. The cation selectivity of gA is mainly
due to these oxygens, which attract cations and repel anions
\cite{jordan90-1133,roux00-13295,kuyucak01-1427}.

In spite of its structural simplicity, the energetics of the ion transport through
gA is far from trivial. Computationally, most of the difficulty arises from the
sensitivity to errors due to finite-size effects and from the poor description of
the polarization effects by the existing force-fields. Besides the cation gA also
accommodates $\sim 6$ single-file water molecules \cite{finkelstein81-155} (see
Fig.~\ref{fig:gram-a}b) whose arrangement and orientation seems to play an
important role in stabilizing the ion within the channel \cite{allen04-117}.

Previous PMF calculations of the potassium ion, $K^{+}$, through gA yielded a
large central barrier that resulted in a conductance orders of magnitude below
those measured. In has been speculated that the measured conductance can be
reproduced by a PMF that has a $\sim 8$~k$_B$T deep energy well at both ends of
the channel and a $\sim 5$~k$_B$T barrier in the middle
\cite{edwards02-1348}. Although PMF calculation methods that try to compensate for
finite-size and polarization effects have improved in recent years, they continue
to yield results that do not match the experimental ones. Most of these methods
employ equilibrium MD simulations with umbrella sampling
\cite{allen06-3447,bastug06-2285,allen04-679} and combined MD simulations with
continuum electrostatics theory \cite{mamonov03-3646}. Recent attempt to apply the
JE method (see Sec.~\ref{sec:theory}) for calculating the PMF of $K^{+}$ in gA did
not yield the desired result \cite{bastug08-155104}. Here we apply our FR method
to calculate both the PMF, $U(z)$, and the position dependent diffusion
coefficient, $D(z)$, of $K^{+}$ in gA, and compare our results with the ones from
the literature.

\subsection{Computer modeling and SMD simulations}
\label{gA-modeling}

The computer model of gA was constructed from its high resolution NMR structure
(Protein Data Bank code 1JNO \cite{townsley01-11676}). After adding the missing
hydrogens, the structure was energy minimized. Using the VMD \cite{humphrey96-33}
plugin \textit{Membrane} the system was inserted into a previously
pre-equilibrated patch of POPE lipid bilayer with size $72\times
72$~\AA$^2$. Lipids within $0.55$~\AA\ of the protein were removed. Then, the
membrane-protein complex was solvated in water, using the VMD plugin
\textit{Solvate}. The final system contained a total of $36,727$ atoms, including
155 lipid molecules and $5,700$ water molecules.
After proper energy minimization and $0.5$~ns long equilibration of the system, a
$K^+$ ion was added at the entrance of the channel. To preserve change neutrality
a $Cl^-$ counterion was also added to the solvent.
Finally, the system was again energy minimized for $10,000$ steps and equilibrated
for $0.5$~ns with $K^{+}$ placed in three different positions along the $z$-axis
of the channel, namely at $z \in \{ -15, 0, 15\}$~\AA. The origin of the $z$-axis
corresponded to the middle of gA (see Fig.~\ref{fig:gram-a}b).  In order to
prevent the pore from being dragged during the SMD pulls of the $K^{+}$ ion, two
types of restraints were imposed: (i) backbone atoms restrained to their
equilibrium positions (referred to as \textit{fully restrained}); and (ii)
backbone atoms restrained only along the $z$-axis (referred to as
\textit{z-restrained}).

The F and R SMD simulations (needed to obtain the PMF using the FR method) were
performed on three systems: (S1) backbone of the channel fully restrained with
only one pair of $K^{+}$ and $Cl^{-}$ ions in the system; (S2) backbone of the
channel fully restrained with $200$~mM electrolyte concentration (obtained by
adding $20$ extra pairs of $K^+$ and $Cl^-$ ions to the solvent using the VMD
plugin \textit{Autoionize}); and (S3) backbone of the channel $z$-restrained and
electrolyte concentration $200$~mM.
A total of $10$ F and $10$ R SMD pulls were performed along the $z$-axis of gA on
two segments: $z \in [-15, 0]$~\AA\ and $z \in [0, 15]$~\AA, corresponding to the
two helical monomers. The pulling speed was $v = 15$~\AA/ns, while the spring
constant of the harmonic potential that guided $K^{+}$ across the pore was $k =
20$~kcal/mol/\AA$^{2}$.

\subsection{Results and discussion}
\label{sec:gA-results}

\begin{figure}[htbp]
   \centering
   \includegraphics[width=3.4in]{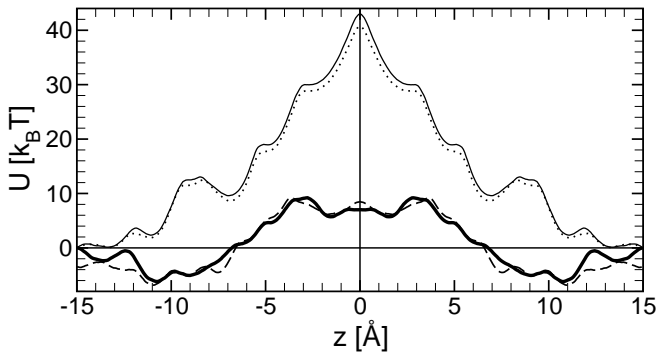} 
   \caption{Comparison of PMFs obtained for systems S1 (thin solid line), S2
     (dotted line) and S3 for two different pulling protocols: (i) pulling force
     on $K^{+}$ applied along the $z$-direction (dashed line), and (ii) $K^{+}$
     pulled along the axis of the channel (thick solid line).}
    \label{fig:pmf-protocol-comp}
\end{figure}

A comparison of the PMFs of $K^{+}$ along the axis of gA obtained for systems S1,
S2 and S3 by employing the FR method is shown in Fig.~\ref{fig:pmf-protocol-comp}.
For gA with fully restrained backbones (i.e., systems S1 and S2) the PMFs have
only a weak dependence on the electrolyte concentration, and exhibit a huge
central potential barrier of $\sim 40$~k$_B$T, which is due to the artificially
imposed rigidity of the system.
Once the flexibility of the gA channel in the plane of the membrane is restored by
restraining the backbone atoms only along the $z$-axis (i.e., system S3), the
central barrier of the PMF decreases to $\sim 15$~k$_B$T, as shown in
Fig.~\ref{fig:pmf-protocol-comp} (thick solid line). The transverse flexibility of
the channel leads to fluctuations in its radius that facilitate the diffusion of
$K^{+}$ along the pore. This is in total agreement with previously published
results, which emphasize the crucial role played by the flexibility of the gA
channel in its cation transport properties
\cite{allen04-679,bastug06-2285,corry05-208}.
The PMF for system S3 was determined with the FR method by employing two different
pulling protocols. First, the pulling force on $K^{+}$ was applied along the
$z$-axis (dashed line in Fig.~\ref{fig:pmf-protocol-comp}) but there was no
restrain on the cation's motion in the cross section of the pore (i.e., in the
$xy$-plane). In the second set of pullings, beside the elastic pulling force
oriented along the $z$-axis, the potassium ion was constrained to move along the
axis of the channel (thick solid line in Fig.~\ref{fig:pmf-protocol-comp}). As one
can see in Fig.~\ref{fig:pmf-protocol-comp}, both pulling protocols yielded
essentially the same PMF.  Thus, we preferred using routinely the second pulling
method especially because during the first one the potassium ion occasionally
escaped between the two helices into the lipid bilayer.

\begin{figure}[htbp]
   \centering
   \includegraphics[width=3.4in]{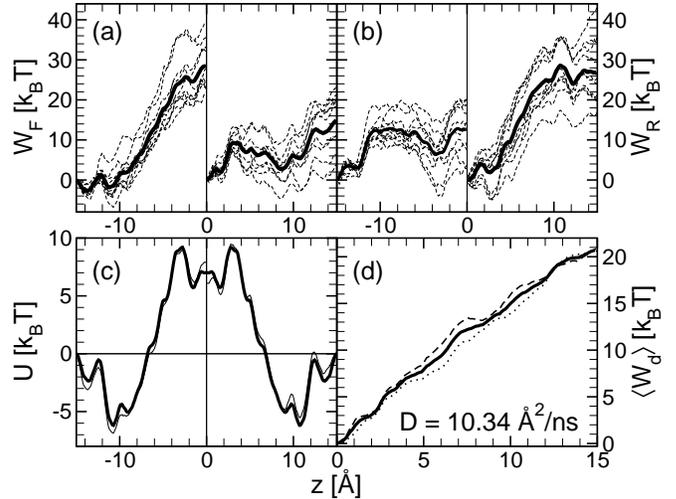} 
   \caption{Results of the FR method calculations for system S3 and SMD pulling
     speed $v = 15$~\AA/ns. (a) individual (thin lines) and mean (thick line) work
     for F pulls; (b) individual (thin lines) and mean (thick line) work for R
     pulls; (c) $U(z)$ along the two segments (dashed line); the symmetrized PMF
     is shown as solid line (see text); (d) mean dissipative work
     $\overline{W}_d(z)$ along the two segments, $z \in (-15, 0)$~\AA\ (dotted
     line) and $z \in (0, 15)$~\AA\ (dashed line), and their arithmetic mean
     (solid line). The slope of the linear $\overline{W}_d(z)$ yields a constant
     diffusion coefficient $D = 10.34$~\AA$^{2}$/ns.}
   \label{fig:work-pmf}
\end{figure}

The PMF, $U(z)$, was calculated separately for the two segments (corresponding to
the two helical monomers) using Eqs.~\eqref{eq:fr-2}. The work done during the F
and R SMD pullings are plotted in Fig.~\ref{fig:work-pmf}a and
Fig.~\ref{fig:work-pmf}b, respectively.
Due to the symmetry of gA with respect to its center, the PMF for the two segments
(dashed lines in Fig.~\ref{fig:work-pmf}c) form nearly mirror-images. Therefore, a
better estimate of the PMF for the entire gA can be obtained by symmetrizing
$U(z)$ with respect to the center of the channel (i.e. $z = 0$~\AA) (solid line in
Fig.~\ref{fig:work-pmf}c). The F and R mean dissipative works,
$\overline{W}_d^{F/R}(z)$, (averaged over the two segments) are also shown in
Fig.~\ref{fig:work-pmf}d (dotted and dashed lines, respectively). The fact that
$\overline{W}_d^F(z)$ and $\overline{W}_d^R(z)$ closely match each other is
another indication that our FR method seems to work fine in the case of the gA
channel too. Note that $\overline{W}_d(z)$, averaged over the F and R processes,
(thick line in Fig.~\ref{fig:work-pmf}d) is almost linear, which according to
Eq.~\eqref{eq:fr-diff} yields a constant diffusion coefficient $D \approx
10.3$~\AA$^2$/ns. Now, the obtained $D$ and $U(z)$ can be used to solve
Eqs.~\eqref{eq:fr-L} and/or \eqref{eq:fr-FP} for making prediction on the long
time dynamics of the $K^{+}$ ion in the gA channel.

\begin{figure}[htbp]
   \centering
   \includegraphics[width=3.4in]{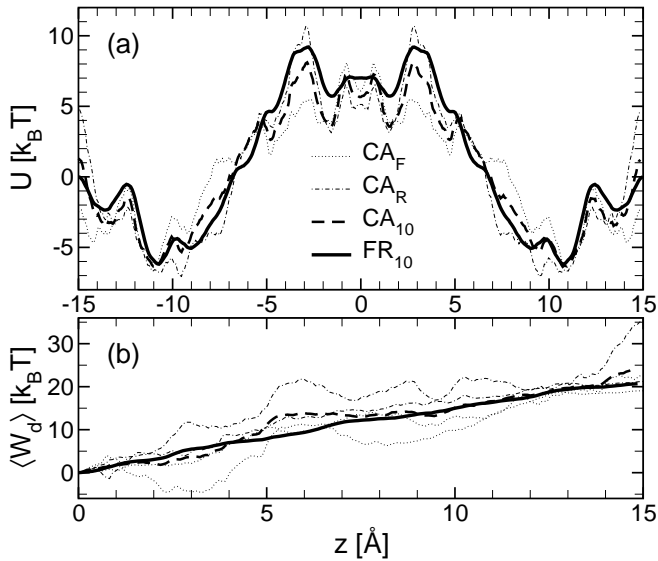} 
   \caption{Comparison between the FR method (thick solid lines) and the cumulant
     approximation of the JE approach (thick dashed lines) for the: (a) potential
     of mean force and (b) mean dissipative work, obtained from a small number
     (only $10$) of F and R fast SMD pulling trajectories. The JE method was
     employed in both F (dotted lines) and R (dash-dotted lines) directions.}
   \label{fig:PMF-Wd_FR-vs-JE}
\end{figure}

The comparison between $U(z)$ and $\overline{W}_d(z)$ obtained from the FR method
(thick solid lines) and the cumulant approximation (CA) of the JE approach (thick
dashed lines), respectively, is shown in Fig.~\ref{fig:PMF-Wd_FR-vs-JE}a. The bias
in the cumulant approximation of the JE method applied either to the F (CA$_F$,
dotted lines) or to the R (CA$_R$, dash-dotted lines) processes is manifest in
Fig.~\ref{fig:PMF-Wd_FR-vs-JE}. While the former (CA$_F$) systematically
underestimates the peaks in the PMF and the corresponding mean dissipative work,
the latter (CA$_R$) systematically overestimates the same quantities. The
difference between the central barrier height of the CA$_F$ and CA$_R$ PMFs is
$3.5$~k$_B$T, while at the channel entrance the difference is almost twice as big
($7$~k$_B$T). The negative (positive) bias in CA$_F$ (CA$_R$) is due to the fact
that the JE approach uses explicitly the variance (i.e., the 2nd cumulant) of the
corresponding non-equilibrium work distributions, which (unlike the mean work)
cannot be accurately estimated from a few SMD pullings (see
Sec.~\ref{sec:theory}). However, by averaging CA$_F$ and CA$_R$ the opposite
biases more or less cancel out and the resulting mean PMF (thick dashed line in
Fig.~\ref{fig:PMF-Wd_FR-vs-JE}a) becomes a close match to $U(z)$ calculated from
the FR method.
According to Fig.~\ref{fig:PMF-Wd_FR-vs-JE}b, the same conclusion can be drawn for
the mean dissipative work as well. 

\begin{figure}[htbp]
   \centering
   \includegraphics[width=3.4in]{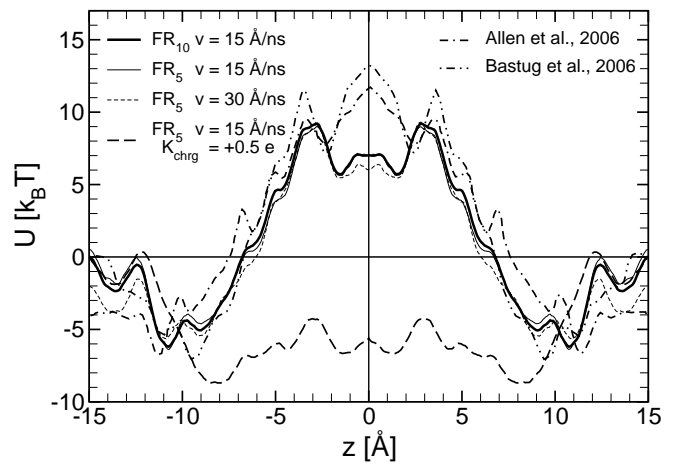} 
   \caption{PMFs of $K^{+}$ in the gA channel determined from the FR method (by
     employing different SMD pulling protocols as described in the text), and the
     umbrella sampling method \cite{bastug06-2285,allen06-3447}.}
   \label{fig:FRvsUS}
\end{figure}

Our $U(z)$, calculated using the FR method, (thick solid line in
Fig.~\ref{fig:FRvsUS}) has two $\sim 6$~$k_BT$ deep wells positioned at the
entrances in the channel ($z \approx \pm 10.8$\AA) and two high barriers of $\sim
15$~$k_BT$ positioned close to the center of the channel ($z \approx \pm
3$~\AA). Another small barrier ($\sim 1.4$~$k_BT$) appears to be located between
the two high barriers, right at the geometrical center of gA. This small center
barrier is well separated by the two main ones by a potential well of $\sim
3.5$~k$_B$T.
According to Fig.~\ref{fig:FRvsUS}, our PMF (thick solid line) is rather similar
to the ones reported in recent publications by Bastug \textit{et al}
\cite{bastug06-2285} (double-dot-dashed line) and by Allen \textit{et al}
\cite{allen06-3447} (dot-dashed line). These authors used the standard umbrella
sampling (US) method \cite{roux95-275,kumar95-1339} to calculate their PMFs. As
shown in Fig.~\ref{fig:FRvsUS}, besides the small difference in the positions of
the wells at the ends of the gA channel, there are two notable differences between
the PMFs obtained by the FR and US methods. First, the barrier height of the PMF
computed with the FR method is only $\sim 15$~k$_B$T as compared to $\sim
20$~k$_B$T obtained from US. Second, the central peak in $U(z)$ obtained from the
FR (US) method is $\sim 2$~k$_B$T below (above) the two main peaks.

To test the reliability of the FR method for determining $U(z)$, besides the
standard pulling protocol (involving $10$ SMD pulls in both F and R directions
with a pulling speed $v=15$~\AA/ns), we have used two additional ones, involving
only $5$ SMD pulls in both F and R directions. The two pulling protocols
differed only in their pulling speeds, namely $v=15$~\AA/ns in the first
(thin-solid line in Fig.~\ref{fig:FRvsUS}) and $v = 30$~\AA/ns (dotted line in
Fig.~\ref{fig:FRvsUS}) in the second. As seen in Fig.~\ref{fig:FRvsUS}, all three
FR method calculations yielded a consistent PMF, with noticeable differences only
around the ends of the gA channel. 

Although the FR method leads to $U(z)$ similar to the US result (albeit with a
smaller main barrier height) none of these PMFs is suitable for reproducing the
experimentally measured $K^{+}$ conductivity of the gA channel. This would require
a channel entrance well depth of $\sim 8$~k$_B$T and a main barrier height of
$\sim 5$~k$_B$T \cite{edwards02-1348}.
The main problems in getting these values are due to the limitations of the
currently used MD methods that use empirical non-polarizable forcefields and,
therefore, cannot account for the induced polarization in the lipid hydrocarbons
and, most importantly, for the polarization of water in the course of the MD
simulations \cite{allen06-3447,allen06-251}.

In order to mimic polarization effects caused by the passage of $K^+$ through the
channel, we reduced the partial charge of the ion from $+e$ to $+0.5e$ in system
S3 (see Sec.~\ref{gA-modeling}), and carried out new SMD F and R pullings for
recalculating the PMF through the FR method. The resulting $U(z)$ is shown in
Fig.~\ref{fig:FRvsUS} (dashed line).  As one can see, in the new PMF the potential
wells at the entrance of the channel moved by $2.5$~\AA\ towards the center and
their depth increased to $8.5$~$k_BT$. Furthermore, in a more dramatic change, the
height of the barrier decreased from $\sim 15$~k$_B$T to $\sim
4.2$~k$_B$T. Although the above approach to account for polarization effects is
rather simplistic, the obtained PMF (apart from the new positions of the potential
wells) has the previously estimated form \cite{edwards02-1348} that is capable for
describing quantitatively the transport of $K^{+}$ in gA.

\section{Conclusions}
\label{sec:conc}

In this paper we have shown that the \textit{FR method} \cite{kosztin06-064106}
provides an effective approach for calculating both the PMF, $U(R)$, and the
diffusion coefficient, $D(R)$, along a properly chosen reaction coordinate $R$, in
biomolecular systems by using only a small number of fast forward and time
reversed constant velocity SMD simulations.
The obtained PMFs for deca-alanine are in good agreement with the ones reported in
recent studies \cite{henin04-2904,park03-3559}. We have found that computationally
the FR method is more efficient and accurate than similar PMF calculation methods,
e.g., the one based on the Jarzynski equality.  By employing the computed PMF and
diffusion coefficient in a suitable stochastic model we could estimate important
characteristics of the studied systems, e.g., the mean folding time of the
stretched deca-alanine.

We also applied the FR method to calculate the PMF of a potassium ion through the
gramicidin A channel. As expected from previous \emph{umbrella sampling}
calculations, the obtained PMF featured a main central barrier of height $\sim
15$~k$_B$T and two wells at the entrance in the channel with depth $\sim
6$~$k_BT$. The PMF was reproduced rather well when using a smaller number of SMD
pulling trajectories and/or higher SMD pulling speeds, confirming the reliability
of the FR method.  The channel protein flexibility, maintained in the SMD
simulations by restraining the corresponding backbone atoms only along the axis of
the channel, has been shown to play a major role in the transport of $K^+$ in
gramicidin A. Indeed, the height of the main potential barrier in a rigid channel
is almost three times higher than in the flexible one. The dissipative work inside
the channel was found to be linear in $z$, yielding a constant diffusion
coefficient $D \approx 10.3$~\AA$^2$/ns.
The PMF calculated from the same SMD pulls using Jarzynski's equality with the
cumulant approximation yielded inconsistent results for both forward and reverse
directions. However, the biases in these to directions almost cancel out when
averaging the forward and reverse PMFs, leading to almost the same result as the
FR method.
Furthermore, the FR method yielded consistently PMFs similar to the ones using the
traditional umbrella sampling method but in considerably less time (i.e., $\sim 3$
days per PMF on a 64~CPU, 2.8GHz Intel Xeon EM64T, cluster).  
However, the conduction of the channel cannot be reproduced with any of the
computed PMF profiles, mainly because of the very large central barrier. The main
problem in determining PMFs in ion channels through MD simulations is the poor
treatment of polarization effects by the current non-polarizable forcefields. To
account for the polarization of K$^+$ inside the channel, its effective point
charge was reduced to $+0.5e$. The recalculated PMF exhibited barrier and well sizes
very close to the values needed to reproduce the experimental data.
Hopefully, with new polarizable force fields the FR method will provide a simple
to use, efficient and reliable tool for calculating PMFs for ion and molecular
transport through channel proteins.

\section*{Acknowledgments}
This work was supported in part by grants from the the Institute for Theoretical
Sciences, a joint institute of Notre Dame University and Argonne National
Laboratory, the U.S. Department of Energy, Office of Science (contract
No.~W-31-109-ENG-38), and the National Science Foundation (FIBR-0526854).
We gratefully acknowledge the generous computational resources provided by the
University of Missouri Bioinformatics Consortium.
M.~Forney gratefully acknowledges a fellowship from the University of Missouri
Undergraduate Research Scholars Program.


\begin{thebibliography}{40}
\expandafter\ifx\csname natexlab\endcsname\relax\def\natexlab#1{#1}\fi
\expandafter\ifx\csname bibnamefont\endcsname\relax
  \def\bibnamefont#1{#1}\fi
\expandafter\ifx\csname bibfnamefont\endcsname\relax
  \def\bibfnamefont#1{#1}\fi
\expandafter\ifx\csname citenamefont\endcsname\relax
  \def\citenamefont#1{#1}\fi
\expandafter\ifx\csname url\endcsname\relax
  \def\url#1{\texttt{#1}}\fi
\expandafter\ifx\csname urlprefix\endcsname\relax\def\urlprefix{URL }\fi
\providecommand{\bibinfo}[2]{#2}
\providecommand{\eprint}[2][]{\url{#2}}

\bibitem[{\citenamefont{Roux}(2002)}]{roux02-182}
\bibinfo{author}{\bibfnamefont{B.}~\bibnamefont{Roux}}, \bibinfo{journal}{Curr.
  Opin. Struct. Biol.} \textbf{\bibinfo{volume}{12}}, \bibinfo{pages}{182}
  (\bibinfo{year}{2002}).

\bibitem[{\citenamefont{Becker et~al.}(2001)\citenamefont{Becker, MacKerell,
  Roux, and Watanabe}}]{CBB01}
\bibinfo{editor}{\bibfnamefont{O.~M.} \bibnamefont{Becker}},
  \bibinfo{editor}{\bibfnamefont{A.~D.} \bibnamefont{MacKerell}},
  \bibinfo{editor}{\bibfnamefont{B.}~\bibnamefont{Roux}}, \bibnamefont{and}
  \bibinfo{editor}{\bibfnamefont{M.}~\bibnamefont{Watanabe}}, eds.,
  \emph{\bibinfo{title}{Computational Biochemistry and Biophysics}}
  (\bibinfo{publisher}{Marcel Dekker}, \bibinfo{address}{New York},
  \bibinfo{year}{2001}).

\bibitem[{\citenamefont{Kosztin et~al.}(2006)\citenamefont{Kosztin, Barz, and
  Janosi}}]{kosztin06-064106}
\bibinfo{author}{\bibfnamefont{I.}~\bibnamefont{Kosztin}},
  \bibinfo{author}{\bibfnamefont{B.}~\bibnamefont{Barz}}, \bibnamefont{and}
  \bibinfo{author}{\bibfnamefont{L.}~\bibnamefont{Janosi}},
  \bibinfo{journal}{J. Chem. Phys.} \textbf{\bibinfo{volume}{124}},
  \bibinfo{pages}{064106} (\bibinfo{year}{2006}).

\bibitem[{\citenamefont{Frenkel and Smit}(2002)}]{frenkel2002}
\bibinfo{author}{\bibfnamefont{D.}~\bibnamefont{Frenkel}} \bibnamefont{and}
  \bibinfo{author}{\bibfnamefont{B.}~\bibnamefont{Smit}},
  \emph{\bibinfo{title}{Understanding Molecular Simulation From Algorithms to
  Applications}} (\bibinfo{publisher}{Academic Press},
  \bibinfo{address}{California}, \bibinfo{year}{2002}).

\bibitem[{\citenamefont{Roux}(1995)}]{roux95-275}
\bibinfo{author}{\bibfnamefont{B.}~\bibnamefont{Roux}},
  \bibinfo{journal}{Comput. Phys. Commun.} \textbf{\bibinfo{volume}{91}},
  \bibinfo{pages}{275} (\bibinfo{year}{1995}).

\bibitem[{\citenamefont{Torrie and Valleau}(1977)}]{torrie77-187}
\bibinfo{author}{\bibnamefont{Torrie}} \bibnamefont{and}
  \bibinfo{author}{\bibnamefont{Valleau}}, \bibinfo{journal}{Journal of
  Computational Physics} \textbf{\bibinfo{volume}{23}}, \bibinfo{pages}{187}
  (\bibinfo{year}{1977}).

\bibitem[{\citenamefont{Park and Schulten}(2004)}]{park04-5946}
\bibinfo{author}{\bibfnamefont{S.}~\bibnamefont{Park}} \bibnamefont{and}
  \bibinfo{author}{\bibfnamefont{K.}~\bibnamefont{Schulten}},
  \bibinfo{journal}{J. Chem. Phys.} \textbf{\bibinfo{volume}{120}},
  \bibinfo{pages}{5946} (\bibinfo{year}{2004}).

\bibitem[{\citenamefont{Park et~al.}(2003)\citenamefont{Park, Khalili-araghi,
  Tajkhorshid, and Schulten}}]{park03-3559}
\bibinfo{author}{\bibfnamefont{S.}~\bibnamefont{Park}},
  \bibinfo{author}{\bibfnamefont{F.}~\bibnamefont{Khalili-araghi}},
  \bibinfo{author}{\bibfnamefont{E.}~\bibnamefont{Tajkhorshid}},
  \bibnamefont{and} \bibinfo{author}{\bibfnamefont{K.}~\bibnamefont{Schulten}},
  \bibinfo{journal}{J. Chem. Phys.} \textbf{\bibinfo{volume}{119}},
  \bibinfo{pages}{3559} (\bibinfo{year}{2003}).

\bibitem[{\citenamefont{Isralewitz et~al.}(2001)\citenamefont{Isralewitz,
  Baudry, Gullingsrud, Kosztin, and Schulten}}]{isralewitz01-13}
\bibinfo{author}{\bibfnamefont{B.}~\bibnamefont{Isralewitz}},
  \bibinfo{author}{\bibfnamefont{J.}~\bibnamefont{Baudry}},
  \bibinfo{author}{\bibfnamefont{J.}~\bibnamefont{Gullingsrud}},
  \bibinfo{author}{\bibfnamefont{D.}~\bibnamefont{Kosztin}}, \bibnamefont{and}
  \bibinfo{author}{\bibfnamefont{K.}~\bibnamefont{Schulten}},
  \bibinfo{journal}{J. Mol. Graph.} \textbf{\bibinfo{volume}{19}},
  \bibinfo{pages}{13} (\bibinfo{year}{2001}).

\bibitem[{\citenamefont{Jensen et~al.}(2002)\citenamefont{Jensen, Park,
  Tajkhorshid, and Schulten}}]{jensen02-6731}
\bibinfo{author}{\bibfnamefont{M.~O.} \bibnamefont{Jensen}},
  \bibinfo{author}{\bibfnamefont{S.}~\bibnamefont{Park}},
  \bibinfo{author}{\bibfnamefont{E.}~\bibnamefont{Tajkhorshid}},
  \bibnamefont{and} \bibinfo{author}{\bibfnamefont{K.}~\bibnamefont{Schulten}},
  \bibinfo{journal}{Proc. Natl. Acad. Sci. U. S. A.}
  \textbf{\bibinfo{volume}{99}}, \bibinfo{pages}{6731} (\bibinfo{year}{2002}).

\bibitem[{\citenamefont{Crooks}(2000)}]{crooks00-2361}
\bibinfo{author}{\bibfnamefont{G.~E.} \bibnamefont{Crooks}},
  \bibinfo{journal}{Phys. Rev. E} \textbf{\bibinfo{volume}{61}},
  \bibinfo{pages}{2361} (\bibinfo{year}{2000}).

\bibitem[{\citenamefont{Hummer and Szabo}(2001)}]{hummer01-3658}
\bibinfo{author}{\bibfnamefont{G.}~\bibnamefont{Hummer}} \bibnamefont{and}
  \bibinfo{author}{\bibfnamefont{A.}~\bibnamefont{Szabo}},
  \bibinfo{journal}{Proc. Natl. Acad. Sci. U. S. A.}
  \textbf{\bibinfo{volume}{98}}, \bibinfo{pages}{3658} (\bibinfo{year}{2001}).

\bibitem[{\citenamefont{Shirts et~al.}(2003)\citenamefont{Shirts, Bair, Hooker,
  and Pande}}]{shirts03-140601}
\bibinfo{author}{\bibfnamefont{M.~R.} \bibnamefont{Shirts}},
  \bibinfo{author}{\bibfnamefont{E.}~\bibnamefont{Bair}},
  \bibinfo{author}{\bibfnamefont{G.}~\bibnamefont{Hooker}}, \bibnamefont{and}
  \bibinfo{author}{\bibfnamefont{V.~S.} \bibnamefont{Pande}},
  \bibinfo{journal}{Phys. Rev. Lett.} \textbf{\bibinfo{volume}{91}},
  \bibinfo{pages}{140601} (\bibinfo{year}{2003}).

\bibitem[{\citenamefont{Zwanzig}(2001)}]{zwanzig2001}
\bibinfo{author}{\bibfnamefont{R.}~\bibnamefont{Zwanzig}},
  \emph{\bibinfo{title}{Nonequilibrium statistical mechanics}}
  (\bibinfo{publisher}{Oxford University Press}, \bibinfo{address}{Oxford ; New
  York}, \bibinfo{year}{2001}), \bibinfo{note}{00023880 Robert Zwanzig.
  Includes bibliographical references (p. 211) and index.}

\bibitem[{\citenamefont{Risken}(1996)}]{risken96}
\bibinfo{author}{\bibfnamefont{H.}~\bibnamefont{Risken}},
  \emph{\bibinfo{title}{The Fokker-Planck Equation: Methods of Solution and
  Applications}} (\bibinfo{publisher}{Springer-Verlag Telos},
  \bibinfo{year}{1996}), \bibinfo{edition}{3rd} ed.

\bibitem[{\citenamefont{Henin and Chipot}(2004)}]{henin04-2904}
\bibinfo{author}{\bibfnamefont{J.}~\bibnamefont{Henin}} \bibnamefont{and}
  \bibinfo{author}{\bibfnamefont{C.}~\bibnamefont{Chipot}},
  \bibinfo{journal}{J. Chem. Phys.} \textbf{\bibinfo{volume}{121}},
  \bibinfo{pages}{2904} (\bibinfo{year}{2004}).

\bibitem[{\citenamefont{Procacci et~al.}(2006)\citenamefont{Procacci, Marsili,
  Barducci, Signorini, and Chelli}}]{procacci06-164101}
\bibinfo{author}{\bibfnamefont{P.}~\bibnamefont{Procacci}},
  \bibinfo{author}{\bibfnamefont{S.}~\bibnamefont{Marsili}},
  \bibinfo{author}{\bibfnamefont{A.}~\bibnamefont{Barducci}},
  \bibinfo{author}{\bibfnamefont{G.~F.} \bibnamefont{Signorini}},
  \bibnamefont{and} \bibinfo{author}{\bibfnamefont{R.}~\bibnamefont{Chelli}},
  \bibinfo{journal}{J Chem Phys} \textbf{\bibinfo{volume}{125}},
  \bibinfo{pages}{164101} (\bibinfo{year}{2006}).

\bibitem[{\citenamefont{Humphrey
  et~al.}(1996{\natexlab{a}})\citenamefont{Humphrey, Dalke, and
  Schulten}}]{HUMP96}
\bibinfo{author}{\bibfnamefont{W.}~\bibnamefont{Humphrey}},
  \bibinfo{author}{\bibfnamefont{A.}~\bibnamefont{Dalke}}, \bibnamefont{and}
  \bibinfo{author}{\bibfnamefont{K.}~\bibnamefont{Schulten}},
  \bibinfo{journal}{Jour. Mol. Graph.} \textbf{\bibinfo{volume}{14}},
  \bibinfo{pages}{33} (\bibinfo{year}{1996}{\natexlab{a}}).

\bibitem[{\citenamefont{Phillips et~al.}(2005)\citenamefont{Phillips, Braun,
  Wang, Gumbart, Tajkhorshid, Villa, Chipot, Skeel, Kale, and
  Schulten}}]{phillips05-1781}
\bibinfo{author}{\bibfnamefont{J.~C.} \bibnamefont{Phillips}},
  \bibinfo{author}{\bibfnamefont{R.}~\bibnamefont{Braun}},
  \bibinfo{author}{\bibfnamefont{W.}~\bibnamefont{Wang}},
  \bibinfo{author}{\bibfnamefont{J.}~\bibnamefont{Gumbart}},
  \bibinfo{author}{\bibfnamefont{E.}~\bibnamefont{Tajkhorshid}},
  \bibinfo{author}{\bibfnamefont{E.}~\bibnamefont{Villa}},
  \bibinfo{author}{\bibfnamefont{C.}~\bibnamefont{Chipot}},
  \bibinfo{author}{\bibfnamefont{R.~D.} \bibnamefont{Skeel}},
  \bibinfo{author}{\bibfnamefont{L.}~\bibnamefont{Kale}}, \bibnamefont{and}
  \bibinfo{author}{\bibfnamefont{K.}~\bibnamefont{Schulten}},
  \bibinfo{journal}{J. Comput. Chem.} \textbf{\bibinfo{volume}{26}},
  \bibinfo{pages}{1781} (\bibinfo{year}{2005}).

\bibitem[{\citenamefont{{MacKerell Jr.} et~al.}(1992)\citenamefont{{MacKerell
  Jr.}, Bashford, Bellott et~al.}}]{MACK92short}
\bibinfo{author}{\bibfnamefont{A.~D.} \bibnamefont{{MacKerell Jr.}}},
  \bibinfo{author}{\bibfnamefont{D.}~\bibnamefont{Bashford}},
  \bibinfo{author}{\bibfnamefont{M.}~\bibnamefont{Bellott}},
  \bibnamefont{et~al.}, \bibinfo{journal}{{FASEB} J.}
  \textbf{\bibinfo{volume}{6}}, \bibinfo{pages}{A143} (\bibinfo{year}{1992}).

\bibitem[{\citenamefont{{MacKerell Jr.} et~al.}(1998)\citenamefont{{MacKerell
  Jr.}, Bashford, Bellott et~al.}}]{MACK98short}
\bibinfo{author}{\bibfnamefont{A.~D.} \bibnamefont{{MacKerell Jr.}}},
  \bibinfo{author}{\bibfnamefont{D.}~\bibnamefont{Bashford}},
  \bibinfo{author}{\bibfnamefont{M.}~\bibnamefont{Bellott}},
  \bibnamefont{et~al.}, \bibinfo{journal}{J. Phys. Chem. B}
  \textbf{\bibinfo{volume}{102}}, \bibinfo{pages}{3586} (\bibinfo{year}{1998}).

\bibitem[{\citenamefont{Miyamoto and Kollman}(1992)}]{miyamoto92-952}
\bibinfo{author}{\bibfnamefont{S.}~\bibnamefont{Miyamoto}} \bibnamefont{and}
  \bibinfo{author}{\bibfnamefont{P.~A.} \bibnamefont{Kollman}},
  \bibinfo{journal}{J.~Comp.\ Chem.} \textbf{\bibinfo{volume}{13}},
  \bibinfo{pages}{952} (\bibinfo{year}{1992}).

\bibitem[{\citenamefont{Wallace}(1986)}]{wallace86-295}
\bibinfo{author}{\bibfnamefont{B.~A.} \bibnamefont{Wallace}},
  \bibinfo{journal}{Biophys. J.} \textbf{\bibinfo{volume}{49}},
  \bibinfo{pages}{295} (\bibinfo{year}{1986}).

\bibitem[{\citenamefont{Tian and Cross}(1999)}]{tian99-1993}
\bibinfo{author}{\bibfnamefont{F.}~\bibnamefont{Tian}} \bibnamefont{and}
  \bibinfo{author}{\bibfnamefont{T.~A.} \bibnamefont{Cross}},
  \bibinfo{journal}{J. Mol. Biol.} \textbf{\bibinfo{volume}{285}},
  \bibinfo{pages}{1993} (\bibinfo{year}{1999}).

\bibitem[{\citenamefont{Jordan}(1990)}]{jordan90-1133}
\bibinfo{author}{\bibfnamefont{P.~C.} \bibnamefont{Jordan}},
  \bibinfo{journal}{Biophys. J.} \textbf{\bibinfo{volume}{58}},
  \bibinfo{pages}{1133} (\bibinfo{year}{1990}).

\bibitem[{\citenamefont{Roux et~al.}(2000)\citenamefont{Roux, Berneche, and
  Im}}]{roux00-13295}
\bibinfo{author}{\bibfnamefont{B.}~\bibnamefont{Roux}},
  \bibinfo{author}{\bibfnamefont{S.}~\bibnamefont{Berneche}}, \bibnamefont{and}
  \bibinfo{author}{\bibfnamefont{W.}~\bibnamefont{Im}},
  \bibinfo{journal}{Biochemistry} \textbf{\bibinfo{volume}{39}},
  \bibinfo{pages}{13295} (\bibinfo{year}{2000}).

\bibitem[{\citenamefont{Kuyucak et~al.}(2001)\citenamefont{Kuyucak, Andersen,
  and Chung}}]{kuyucak01-1427}
\bibinfo{author}{\bibfnamefont{S.}~\bibnamefont{Kuyucak}},
  \bibinfo{author}{\bibfnamefont{O.~S.} \bibnamefont{Andersen}},
  \bibnamefont{and} \bibinfo{author}{\bibfnamefont{S.~H.} \bibnamefont{Chung}},
  \bibinfo{journal}{Rep. Prog. Phys.} \textbf{\bibinfo{volume}{64}},
  \bibinfo{pages}{1427} (\bibinfo{year}{2001}).

\bibitem[{\citenamefont{Finkelstein and Andersen}(1981)}]{finkelstein81-155}
\bibinfo{author}{\bibfnamefont{A.}~\bibnamefont{Finkelstein}} \bibnamefont{and}
  \bibinfo{author}{\bibfnamefont{O.~S.} \bibnamefont{Andersen}},
  \bibinfo{journal}{J. Membr. Biol.} \textbf{\bibinfo{volume}{59}},
  \bibinfo{pages}{155} (\bibinfo{year}{1981}).

\bibitem[{\citenamefont{Allen et~al.}(2004{\natexlab{a}})\citenamefont{Allen,
  Andersen, and Roux}}]{allen04-117}
\bibinfo{author}{\bibfnamefont{T.~W.} \bibnamefont{Allen}},
  \bibinfo{author}{\bibfnamefont{O.~S.} \bibnamefont{Andersen}},
  \bibnamefont{and} \bibinfo{author}{\bibfnamefont{B.}~\bibnamefont{Roux}},
  \bibinfo{journal}{Proc. Natl. Acad. Sci. U.S.A.}
  \textbf{\bibinfo{volume}{101}}, \bibinfo{pages}{117}
  (\bibinfo{year}{2004}{\natexlab{a}}).

\bibitem[{\citenamefont{Edwards et~al.}(2002)\citenamefont{Edwards, Corry,
  Kuyucak, and Chung}}]{edwards02-1348}
\bibinfo{author}{\bibfnamefont{S.}~\bibnamefont{Edwards}},
  \bibinfo{author}{\bibfnamefont{B.}~\bibnamefont{Corry}},
  \bibinfo{author}{\bibfnamefont{S.}~\bibnamefont{Kuyucak}}, \bibnamefont{and}
  \bibinfo{author}{\bibfnamefont{S.~H.} \bibnamefont{Chung}},
  \bibinfo{journal}{Biophys. J.} \textbf{\bibinfo{volume}{83}},
  \bibinfo{pages}{1348} (\bibinfo{year}{2002}).

\bibitem[{\citenamefont{Allen et~al.}(2006{\natexlab{a}})\citenamefont{Allen,
  Andersen, and Roux}}]{allen06-3447}
\bibinfo{author}{\bibfnamefont{T.~W.} \bibnamefont{Allen}},
  \bibinfo{author}{\bibfnamefont{O.~S.} \bibnamefont{Andersen}},
  \bibnamefont{and} \bibinfo{author}{\bibfnamefont{B.}~\bibnamefont{Roux}},
  \bibinfo{journal}{Biophys. J.} \textbf{\bibinfo{volume}{90}},
  \bibinfo{pages}{3447} (\bibinfo{year}{2006}{\natexlab{a}}).

\bibitem[{\citenamefont{Bastug et~al.}(2006)\citenamefont{Bastug, Gray-Weale,
  Patra, and S.}}]{bastug06-2285}
\bibinfo{author}{\bibfnamefont{T.}~\bibnamefont{Bastug}},
  \bibinfo{author}{\bibfnamefont{A.}~\bibnamefont{Gray-Weale}},
  \bibinfo{author}{\bibfnamefont{S.~M.} \bibnamefont{Patra}}, \bibnamefont{and}
  \bibinfo{author}{\bibfnamefont{K.}~\bibnamefont{S.}},
  \bibinfo{journal}{Biophys. J.} \textbf{\bibinfo{volume}{90}},
  \bibinfo{pages}{2285} (\bibinfo{year}{2006}).

\bibitem[{\citenamefont{Allen et~al.}(2004{\natexlab{b}})\citenamefont{Allen,
  Andersen, and Roux}}]{allen04-679}
\bibinfo{author}{\bibfnamefont{T.~W.} \bibnamefont{Allen}},
  \bibinfo{author}{\bibfnamefont{O.~S.} \bibnamefont{Andersen}},
  \bibnamefont{and} \bibinfo{author}{\bibfnamefont{B.}~\bibnamefont{Roux}},
  \bibinfo{journal}{J. Gen. Physiol.} \textbf{\bibinfo{volume}{124}},
  \bibinfo{pages}{679} (\bibinfo{year}{2004}{\natexlab{b}}).

\bibitem[{\citenamefont{Mamonov et~al.}(2003)\citenamefont{Mamonov, Coalson,
  Nitzan, and Kurnikova}}]{mamonov03-3646}
\bibinfo{author}{\bibfnamefont{A.~B.} \bibnamefont{Mamonov}},
  \bibinfo{author}{\bibfnamefont{R.~D.} \bibnamefont{Coalson}},
  \bibinfo{author}{\bibfnamefont{A.}~\bibnamefont{Nitzan}}, \bibnamefont{and}
  \bibinfo{author}{\bibfnamefont{M.~G.} \bibnamefont{Kurnikova}},
  \bibinfo{journal}{Biophys. J.} \textbf{\bibinfo{volume}{84}},
  \bibinfo{pages}{3646} (\bibinfo{year}{2003}).

\bibitem[{\citenamefont{Bastug et~al.}({2008})\citenamefont{Bastug, Chen,
  Patra, and Kuyucak}}]{bastug08-155104}
\bibinfo{author}{\bibfnamefont{T.}~\bibnamefont{Bastug}},
  \bibinfo{author}{\bibfnamefont{P.-C.} \bibnamefont{Chen}},
  \bibinfo{author}{\bibfnamefont{S.~M.} \bibnamefont{Patra}}, \bibnamefont{and}
  \bibinfo{author}{\bibfnamefont{S.}~\bibnamefont{Kuyucak}},
  \bibinfo{journal}{{J. Chem. Phys.}} \textbf{\bibinfo{volume}{{128}}},
  \bibinfo{pages}{155104} (\bibinfo{year}{{2008}}).

\bibitem[{\citenamefont{Townsley et~al.}(2001)\citenamefont{Townsley, Tucker,
  Sham, and Hinton}}]{townsley01-11676}
\bibinfo{author}{\bibfnamefont{L.~E.} \bibnamefont{Townsley}},
  \bibinfo{author}{\bibfnamefont{W.~A.} \bibnamefont{Tucker}},
  \bibinfo{author}{\bibfnamefont{S.}~\bibnamefont{Sham}}, \bibnamefont{and}
  \bibinfo{author}{\bibfnamefont{J.~F.} \bibnamefont{Hinton}},
  \bibinfo{journal}{Biochemistry} \textbf{\bibinfo{volume}{40}},
  \bibinfo{pages}{11676} (\bibinfo{year}{2001}).

\bibitem[{\citenamefont{Humphrey
  et~al.}(1996{\natexlab{b}})\citenamefont{Humphrey, Dalke, and
  Schulten}}]{humphrey96-33}
\bibinfo{author}{\bibfnamefont{W.}~\bibnamefont{Humphrey}},
  \bibinfo{author}{\bibfnamefont{A.}~\bibnamefont{Dalke}}, \bibnamefont{and}
  \bibinfo{author}{\bibfnamefont{K.}~\bibnamefont{Schulten}},
  \bibinfo{journal}{J. Mol. Graphics} \textbf{\bibinfo{volume}{14}},
  \bibinfo{pages}{33} (\bibinfo{year}{1996}{\natexlab{b}}).

\bibitem[{\citenamefont{Corry and Chung}(2005)}]{corry05-208}
\bibinfo{author}{\bibfnamefont{B.}~\bibnamefont{Corry}} \bibnamefont{and}
  \bibinfo{author}{\bibfnamefont{S.~H.} \bibnamefont{Chung}},
  \bibinfo{journal}{Eur. Biophys. J.} \textbf{\bibinfo{volume}{34}},
  \bibinfo{pages}{208} (\bibinfo{year}{2005}).

\bibitem[{\citenamefont{Kumar et~al.}(1995)\citenamefont{Kumar, Rosenberg,
  Bouzida, Swendsen, and Kollman}}]{kumar95-1339}
\bibinfo{author}{\bibfnamefont{S.}~\bibnamefont{Kumar}},
  \bibinfo{author}{\bibfnamefont{J.~M.} \bibnamefont{Rosenberg}},
  \bibinfo{author}{\bibfnamefont{D.}~\bibnamefont{Bouzida}},
  \bibinfo{author}{\bibfnamefont{R.~H.} \bibnamefont{Swendsen}},
  \bibnamefont{and} \bibinfo{author}{\bibfnamefont{P.~A.}
  \bibnamefont{Kollman}}, \bibinfo{journal}{J. Comput. Chem.}
  \textbf{\bibinfo{volume}{16}}, \bibinfo{pages}{1339} (\bibinfo{year}{1995}).

\bibitem[{\citenamefont{Allen et~al.}(2006{\natexlab{b}})\citenamefont{Allen,
  Andersen, and Roux}}]{allen06-251}
\bibinfo{author}{\bibfnamefont{T.~W.} \bibnamefont{Allen}},
  \bibinfo{author}{\bibfnamefont{O.~S.} \bibnamefont{Andersen}},
  \bibnamefont{and} \bibinfo{author}{\bibfnamefont{B.}~\bibnamefont{Roux}},
  \bibinfo{journal}{Biophys. Chem.} \textbf{\bibinfo{volume}{124}},
  \bibinfo{pages}{251} (\bibinfo{year}{2006}{\natexlab{b}}).

\end{thebibliography}

\end{document}